%% file: main.tex
\documentclass{article}
\usepackage{spconf,amsmath,graphicx}
\usepackage{amsfonts}
\usepackage{bm}
\usepackage{algorithmic}
\usepackage{algorithm}
\usepackage{array}
\usepackage[caption=false,font=normalsize,labelfont=sf,textfont=sf]{subfig}
\usepackage{textcomp}
\usepackage{stfloats}
\usepackage{url}
\usepackage{verbatim}
\usepackage{cite}
\usepackage{multirow,tabularx,enumitem,makecell,booktabs}
\usepackage{lipsum}
\usepackage{arydshln}
\usepackage{breqn}

\title{Team HYU ASML ROBOVOX SP Cup 2024 System Description}
\name{Jeong-Hwan Choi$^\dag$, Gaeun Kim$^*$, Hee-Jae Lee$^*$, Seyun Ahn$^*$, Hyun-Soo Kim$^*$, and Joon-Hyuk Chang$^\ddag$}
\address{Department of Electronic Engineering \\ Hanyang University, Seoul, Republic of Korea} 
\begin{document}
%
%
\maketitle
\def\thefootnote{\ddag}\footnotetext{Team supervisor and corresponding author.}\def\thefootnote{\arabic{footnote}}
\def\thefootnote{\dag}\footnotetext{Postgraduate student (tutor) and first author.}\def\thefootnote{\arabic{footnote}}
\def\thefootnote{$*$}\footnotetext{Four of them were undergraduate students when we entered the SP Cup 2024. They were attending in Hankuk University of Foreign Studies (Gaeun Kim), Suwon University (Hee-Jae Lee), and Hanyang University (Seyun Ahn and Hyun-Soo Kim), respectively.}\def\thefootnote{\arabic{footnote}}
\begin{abstract}
This report describes the submission of HYU ASML team to the IEEE Signal Processing Cup 2024 (SP Cup 2024). 
This challenge, titled ``ROBOVOX: Far-Field Speaker Recognition by a Mobile Robot,'' focuses on speaker recognition using a mobile robot in noisy and reverberant conditions.
Our solution combines the result of deep residual neural networks and time-delay neural network-based speaker embedding models.
These models were trained on a diverse dataset that includes French speech.
To account for the challenging evaluation environment characterized by high noise, reverberation, and short speech conditions, we focused on data augmentation and training speech duration for the speaker embedding model.
Our submission achieved second place on the SP Cup 2024 public leaderboard, with a detection cost function of $0.5245$ and an equal error rate of $6.46\%$.
\end{abstract}

\section{Introduction}
\label{sec:Introduction}
Advances in speaker embedding models using deep learning have improved speaker recognition performance.
The number and quality of training data are important in the speaker classification capability of speaker embedding.
In particular, in order to learn a speaker embedding model that performs well under specific evaluation conditions, it is necessary to construct a training environment similar to the evaluation conditions. 
In this paper, we present a detailed description of our solution for the IEEE Signal Processing Cup 2024 (SP Cup 2024).
The subtitle of the SP Cup 2024 is ``ROBOVOX: Far-Field Speaker Recognition by a Mobile Robot''. 
This challenge concerns doing far-field text-independent speaker verification (SV) recorded by a mobile robot in indoor conditions.
Therefore, the recorded multi-channel audio was subjected to severe reverberation and included the various noises of opening and closing doors, robot engine sounds, \textit{etc}, that can occur indoors.
The evaluation is done with a single channel, while the enrollment and test utterances are French speech obtained from different channels.
The enrollment is made up of several utterances, while the test is configured with single and short utterances, which makes the challenge harder.
We built the training data and environment considering this specific domain knowledge.
We used ECAPA-TDNN \cite{ecapan-tdnn} and ResNet \cite{resnet, resnet_asv}, which represent speaker embedding models based on 1-dimensional convolution neural network (Conv1D) and Conv2D.
We calculated the score of the test trials using the cosine distance between speaker embeddings and additionally reported the improved performance by utilizing score calibration methods.
Our work with the python and PyTorch toolkit is available on GitHub \footnote{https://github.com/Ribotot/asml-asv}.
\section{Dataset}
\subsection{Train data}
\input{tables/data_stats}
We utilized diverse datasets to train our speaker embedding model, and we summarized them in Table 1.
Further details regarding the additional preprocessing of each dataset are provided below.
Additionally, we have included information on the datasets used for evaluation and data augmentation purposes.
\subsubsection{VoxCeleb2 \cite{vox2}}
The VoxCeleb2 development set (\textit{dev} set) was utilized in training the speaker embedding models.
VoxCeleb2 devset contains 1,092,009 utterances from 5,994 celebrities, extracted from interview videos of celebrity uploaded to YouTube.
This large data is a freely available and has been used in many text-independent SV studies \cite{ecapan-tdnn, resnet_asv2, mfa_conformer, skn_asv, thin_resnet_asv, bc_cmt, short_asv, dino_cho, dino_dyn, semi_dino_asv}.
\subsubsection{CN-Celeb1 \cite{cnceleb1}}
The CN-Celeb1 dataset consists of 126,532 utterances from 997 Chinese celebrities, covering 11 different genres, such as interview, entertainment, \textit{etc}.
The total duration of audio sample is 274 hours and we used all the data used for training and evaluation.
\subsubsection{Short-duration SV (SdSV) 2020 Task 2 dataset \cite{sdsv2020}}
The SdSV 2020 Task 2 dataset is an evaluation for text-independent SV.
It is based on the DeepMine dataset, which was collected using an Android application through crowd-sourcing \cite{deepmine}.
The SdSV 2020 Task 2 dataset consists of 124,903 utterances from Persian native speakers and several English non-native speakers.
For training, we specifically utilized 588 speakers and 85,764 utterances of Persian speech from the SdSV 2020 Task 2 dataset.
\subsubsection{Multilingual TEDx (mTEDx) Corpus \cite{mtedx}}
The source for mTEDx corpus is a multilingual set of TEDx talks, which contains 765 h of audio data in 8 languages.
We only used 189 h of French data. 
The audio data consists of two-channel signals sampled at either 44.1 or 48 kHz, thus we downsampled the first channel signal to 16 kHz.
We classified the data based on speaker identities provided in the description, and we also segmented the audio by utterance. 
Additionally, we combined shorter signals (less than 4 s) from the same speaker, ensuring that each audio sample was longer than 4 s.
The number of speakers and audio sample we got from the pre-processing were 961 and 83,879, respectively.
\subsubsection{Multilingual LibriSpeech (MLS) Corpus \cite{mls}}
The French portion of the MLS Corpus, a dataset derived from audiobooks read and available on LibriVox, was used. It includes about 1.3K hours of speech data from 224 books read by 114 different speakers and is appropriately partitioned into train, development and test sets. The total duration of the training set is about 1K hours, while both the development and test sets are approximately 10 hours each.
\subsubsection{ROBOVOX sample \cite{sp_cup24}}
The challenge provided the two speaker sample files for system adaptation or data augmentation purposes.
We listened to each sample and segmented them based on utterances.
The samples have 8 channels, but we only utilized data from the 4th and 5th channels, which are designated for enrollment and testing, respectively.
Additionally, short utterances were concatenated similar to the preprocessing of the mTEDx Corpus.
\subsection{Validation data}
\subsubsection{VoxCeleb1 \cite{vox1}}
We used the VoxCeleb1 test set for speaker embedding model validation.
It contains 4,874 utterances from 40 speakers and it also provide original (clean) trials. 
We checked the SV performance of the models using the VoxCeleb1 test protocol at every epoch and selected the best-performing epoch's model.
\subsubsection{ROBOVOX Far-field multi-channel tracks \cite{sp_cup24}}
For validation, we utilized additional trials from the far-field multi-channel tracks.
As shown in the challenge evaluation plan \cite{sp_cup24}, the trials in the single-channel track used a 4th channel for enrollment and a 5th channel for testing; similarly, we modified the multi-channel trials to be like the trials in the single-channel track.
We want to be clear that we only used it to verify and report on the SV performance of the model,  and we did not utilize this dataset and trial for any additional fine-tuning or back-end processes.
\subsection{Augmentation data}
To implement a training environment similar to the evaluation environment, we utilized room impulse response (RIR) and noise samples for data augmentation.
During training, the different noisy and reverberated audio samples were randomly conducted each time, as the augmentation was performed online.
We compute the probability of applying each augmentation technique every time. The probabilities for applying reverberation and adding noise are determined independently using different values.
Specifically, we utilized the OpenSLR RIR and Noise corpora \cite{rir} and the MUSAN corpus \cite{musan}.
The OpenSLR RIR and Noise corpora contain 16,200 medium-sized room RIRs and 5,400 large-sized room RIRs.
The MUSAN corpus contains three types of noise: 60 hours of babble, 42 hours of music, and 6 hours of other background noise. 
We also extracted a silent segment from each ROBOVOX sample to serve as additive noise.
In our experiments, each noise type was randomly selected and added to the input audio with an SNR value in the following ranges: MUSAN babble (13$-$20dB SNR of 3$-$7 speeches), music (5$-$15dB SNR), noise (-3$-$15dB SNR), and ROBOVOX sample noise (-3$-$15dB SNR).
\section{Speaker embedding model}
We implemented a large-scale ECAPA-TDNN and ResNet-based speaker embedding model to report the final result.
Before training both models, we conducted experiments using a small-scale speaker embedding model, ECAPA-TDNN-512, to find the appropriate input utterance duration and data augmentation method. 
\subsection{Experiment of ECAPA-TDNN-512 for setting training details}
The smallest model with $6.2$M parameters proposed in the original paper \cite{ecapan-tdnn} of ECAPA-TDNN is ECAPA-TDNN-512.
We used 80-dimensional log Mel-spectrograms as input to ECAPA-TDNN-512, and they were extracted using a 25-ms Hamming window and 10-ms hop size. 
We extracted acoustic features without extra VAD, and the speaker embedding size was $192$.
When the base channel size of ECAPA-TDNN model is C (ECAPA-TDNN-C), the model structure is detailed in Table 2.
\input{tables/ecapa-tdnn-c}
To find the fine training setup by using ECAPA-TDNN-512, we utilized VoxCeleb2, MUSAN, and OpenSLR RIR datasets.
In these experiments, we added the MUSAN noise to the input audio at each trial (with an augmentation probability of 1 for the noise).
On the other hand, the probability of applying an RIR was 0.5.
We employed the Adam optimizer \cite{adam} with a mini-batch size of $256$.
The speaker classification loss was computed using the angular additive margin (AAM) softmax function \cite{arcface} with a margin of $0.2$ and scale of $30$.
A total of 80 epochs were set for training with warm-up cosine scheduling.
The learning rate was linearly increased from $0$ to $10^{-3}$ during the first ten epochs and then decreased to $10^{-6}$.
This training methodology is based on our previous study \cite{voxsrc22_hyu, sasv_hyu22, bcres2net-asc, sdsv20_hyu}. 
Please refer to these studies for more detailed information.
In our experiments with ECAPA-TDNN-512, we utilized the cosine distance between speaker embeddings for scoring. For performance measurement, we employed the equal error rate (EER) and the detection cost function (DCF) as outlined in challenge \cite{sp_cup24}.
The DCF quantifies the significance of minimizing false rejections compared to false acceptances, giving more weight to the former.
The two types of DCF, Day and Night, were used in the challenge. 
The Day DCF were computed with the following parameters: false rejection rate $C_{miss} = 1$, false acceptance rate $C_{fa} = 20$, and prior probability of target speaker occurrence $P_{Target} = 0.8$.
In the case of the Night DCF: false rejection rate $C_{miss} = 10$, false acceptance rate $C_{fa} = 100$, and prior probability of target speaker occurrence $P_{Target} = 0.01$.
We used a new metric of the average of the two DCFs, which we call DCF$_c$, for determination of the final ranking.
Since enrollment takes place over multiple utterances, the prototype embedding was calculated for each speaker by averaging the speaker embeddings from the utterances.
We observed a significant number of silent segments in each ROBOVOX audio sample.
Consequently, we hypothesized that adding noise to the speech would be more effective when its duration  lasted longer than that of the speech.
To validate this hypothesis, we conducted investigations to identify the optimal duration of noise and speech as detailed in Table 3.
Through the experiment, it was confirmed that incorporating silent segments, characterized by the presence of noise without any vocal elements, into the speech is effective for the challenge dataset.
Specifically, the optimal configuration, achieving the lowest EER and DCF$_{c}$, involved an input utterance of 1.8 seconds, accompanied by non-vocal noise (silent segments) that lasted for a total duration of 2.4 seconds, indicating that the noise extends 0.6 seconds beyond the utterance.
An ablation study on configuring the hyperparameters for data augmentation was performed with 2 s of speech and 3 s of noise.
\input{tables/exp_duration}
\input{tables/exp_augmentation}
As presented in Table 4, we observed an enhancement in SV performance when the MUSAN dataset was used without the inclusion of music noise.
This improvement could be attributed to the absence of music noise in the environment of the evaluation database.
When we decreased the probability of applying RIR from 0.5 to 0.25 or increased it to 0.75, both changes resulted in subpar performance.
Furthermore, we identified the occurrence of audio clipping in ROBOVOX sample, and utilized it as an additional data augmentation whether or not the other methods,  reverberation and noise, were introduced.
The clipping augmentation was implemented by limiting the amplitude to $N\%$ of the speech's maximum amplitude, where $N$ was randomly selected from a range of 3 to 8, whenever the amplitude exceeded this threshold.
We found that setting the clipping probability to 0.25 resulted in better outcomes compared to the probability of 0.5.
Based on these experiments, the following subsections describe the structure and training details of the  large-scale speaker embedding model that used to produce the final results.
\subsection{Acoustic feature}
For training on large models, we used 96-dimensional log Mel-spectrograms with a higher resolution of the frequency bins than in our experiments with ECAPA-TDNN-512.
We did not use VADs and augmented the data with the best settings from our previous subsection.
In addition, we applied SpecAug \cite{specaug} by randomly masking the 5 frequency bins and 5 time frames of each input acoustic feature.
\subsection{Speaker embedding models architecture}
The large-scale ECAPA-TDNN and ResNet has achieved state-of-the-art performance in SV field within recent years \cite{ID_voxsrc23, microsoft_voxsrc22, kriston_voxsrc22, spkin_ffsvc2022, voxsrc22_hyu, sasv_hyu22, spkin_2021, id_voxsrc20}.
We used ECAPA-TDNN-2048 and an extend version of ResNet. 
Our ResNet is denoted ResNet-64 because the number of CNNs was set to 64.
ResNet-64 has a kernel size similar to the ID R\&D system \cite{ID_voxsrc23} and two types of pooling layer. 
The first is the channel-dependent attentive statistics pooling (CD-ASP) used in ECAPA-TDNN, and the second is the frequency-statistics-dependent attentive statistics pooling (FS-ASP) used in our previous work \cite{bc_cmt}.
FS-ASP is an extension of the CD-ASP for the frequency axis.
\input{tables/resnet-64}
%
The overall architecture of the ResNet-64 is shown in Table 5.
Note that the speaker embedding size for both ECAPA-TDNN-2048 and ResNet-64 is $256$, and the number of parameters is 45.30M and 33.55M respectively. 
\subsection{Training details}
Our strategy was pre-training with the training datasets described in Section 2 and fine-tuning with the selected datasets.
Each phase was trained with 80 epochs.
In pre-training, we fixed the learning rate at 0.001 for the first 40 epochs and halved it every 5 epochs thereafter. 
The ResNet-64 was trained on four NVIDIA GeForce RTX 4090 GPUs with a batch size of 160, while the ECAPA-TDNN-2048 was trained on a single NVIDIA GeForce RTX 3090 GPU with a batch size of 128.
Loss functions are critical to training the speaker embedding model.
We applied the sub-center top-k AAM softmax loss \cite{spkin_2021, topk} in the pre-training phase and discarded the sub-center and top-k methods in the fine-tuning phase. 
During the pre-training, the sub-center number was set to 3 and the margin was set to increase linearly from 0 to 0.2 in the first 20 epochs.
Furthermore, the penalty and Top-K were set to 0.06 and 5, respectively.
In the fine-tuning phase, we fixed the margin value to 0.2.
The initial learning rate was set to 2e-5 and annealing was the same as pre-training. 
We chose the VoxCeleb2, mTEDx, MLS, and ROBOVOX samples as fine-tuning datasets.
For each dataset, the data augmentation proceeded as follows.
\begin{itemize}[leftmargin=0.25in]
\item We applied reverberation to VoxCeleb2 with a probability of 0.5. Additionally, a sample noise from ROBOVOX and MUSAN's background noise and babble were added.
\item Reverberation and clipping with a probability of 0.5 and 0.25, respectively, were applied to MLS, and ROBOVOX sample noise was added as additional noise.
\item Since mTEDx is mainly recorded in auditoriums and concert halls and contains a lot of reverberation, we applied clipping with a probability of 0.25 and additional noise using the ROBOVOX sample noise.
\item The ROBOVOX sample was used without any data augmentation.
\end{itemize}
The SpecAug was not used in the fine-tuning phase.
\section{Scoring and back-end}
After training, we obtained the weights from the four epochs in a specific order based on the VoxCeleb1 original trials performance. 
We averaged these weights for the final speaker embedding model.
Before arriving at the final result, we investigated additional pre-processing, dereverberation and voice activity detection (VAD), that could improve the SV performance.
We first found that both pre-processing methods introduce SV performance penalties when applied to both enrollment and test utterances.
Our pre-processing methods were only applied to enrollment utterances.
In particular, we used virtual acoustic channel expansion weighted prediction error (VACE-WPE) \cite{vace_wpe_interspeech, vace_taslp}, a single-channel dereverberation algorithm that generates a virtual channel to support eliminate reverberation.
The pre-trained VACE model, specifically optimized for SV task, \cite{tso_vace_taslp}, was available on GitHub \footnote{text{https://github.com/dreadbird06/tso\_vace\_wpe}}.
For VAD. We utilized two pre-trained VAD models \footnote{https://huggingface.co/speechbrain/vad-crdnn-libriparty} \footnote{https://huggingface.co/pyannote/voice-activity-detection} provided by \textit{pyannote.audio 2.0} \cite{pyannote} and \textit{speechbrain} \cite{speechbrain} to detect speech segments. 
We applied VAD models to the original and dereverberated signals, resulting in a total of four VAD results.
For simplicity, we referred to them as VAD$_{pn}$, VAD$_{sb}$, VAD$_{pn-d}$ and VAD$_{sb-d}$ respectively.
\input{fig1}
Our overall pre-processing process is illustrated in Fig.\,1. 
We used the obtained VADs and the dereverberated signals to observe the SV performance, as shown in Tables 6 and 7.
\input{tables/exp_preprocessiong1}
\input{tables/exp_preprocessiong2}
In this experiments, we obtained speaker embedding and VAD for each utterance for the enrollment speakers and scoring phase. 
Because enrollment speakers had multiple assigned utterances, we performed a weighted summation of the speaker embeddings of the enrollment utterances to obtain a prototype of each speaker.
These prototype embedding represented the enrollment speaker.
The weights used to obtain the prototype embedding were the total frame duration of each utterance, leveraging information from the VAD.
Specifically, the weights were computed by dividing the duration of each utterance by the total duration of the utterances to increase the contribution of speaker embeddings with longer utterances.
We also considered the situation where VAD is not detected.
We set the duration of utterance which has no VAD detection to 1e-6. 
Tables 6 and 7 also show whether the VAD was used to remove silent segments or only to calculate the speaker prototype of the enrollment.
In these tables, "segmentation" is unchecked to indicate that the VAD is only used for prototype embedding computations, and checked to indicate that the actual output of the VAD is used after further segmentation.
We observed significant SV performance gains on ECAPA-TDNN-2048 and ResNet-64 when segmenting using VADs.
The SV performance improvement was larger for ECAPA-TDNN-2048 than for ResNet-64.
We also confirmed that dereverberation through the VACE method improves the VAD performance.
Although the enrollment utterance is relatively clean compared to the test utterance, it also contains some reverberant components, thus we can analyze the positive impact of reverberation removal on SV performance.
Based on the results obtained in Tables 6 and 7, we verified the SV performance of the evaluation data under effective pre-processing conditions.
We then used an additional back-end process, including score normalization, quality-aware score calibration, and score fusion to improve SV performance.
\input{tables/exp_final.tex}
\subsection{Score normalization}
The adaptive score normalization (AS-Norm) method in SV effectively reduces domain differences between the training and test datasets.
P. Matějka, \textit{et al.} \cite{socre_norm2} investigated various AS-Norm methods and suggested the speaker-wise AS-Norm method to be effective, and it has been used in many systems of SV challenge \cite{ID_voxsrc23, microsoft_voxsrc22, kriston_voxsrc22, spkin_ffsvc2022, voxsrc22_hyu, sasv_hyu22, spkin_2021, id_voxsrc20}.
We selected the dataset used in fine-tuning without any augmentation to perform the speaker-wise AS-Norm.
Speaker embeddings from these datasets were averaged speaker-wise. 
We then selected the top 400 highest scores to calculate the mean and standard deviation (STD).
When the score for trials is $S$, then the normalized score $S_{as}$, which can be formulated as follows:
\begin{equation}
\label{eqn:as_norm}
    S_{as} =\frac{1}{2}\left(\frac{S-\text{mean}_{e}}{\text{std}_{e}} + \frac{S-\text{mean}_{t}}{\text{std}_{t}}\right),
\end{equation}
where mean$_{e}$, mean$_{t}$, std$_{e}$ and std$_{t}$ denote the mean and STD values for the enrollment and test utterances, respectively.
In this paper, we proposed a novel trainable AS-Norm (TAS-Norm) method for this challenge.
We believed that each utterance's mean and STD were not optimized to normalize the score.
We leveraged these mean and STD obtained from AS-Norm to learn the optimized mean and STD for score normalization.
The normalized score in TAS-Norm is computed as follows:
%
\begin{dmath}
\label{eqn:proposed_as_norm}
    S_{tas} = \frac{1}{2}\left(\frac{S-\sigma\left(\text{w}_{mean}\, \text{mean}_{e}+\text{b}_{mean}\right)}{\sigma\left(\text{w}_{std}\,\text{std}_{e}+\text{b}_{std}\right)} \\ + \frac{S-\sigma\left(\text{w}_{mean}\,\text{mean}_{t}+\text{b}_{mean}\right)}{\sigma\left(\text{w}_{std}\,\text{std}_{t}+\text{b}_{std}\right)} \right),
\end{dmath}
%
where $w_{mean}$, $w_{std}$, $b_{mean}$ and $b_{std}$ are the scalars that we need to learn.
The $\sigma(\cdot)$ denotes the sigmoid activation function. 
These scalars and nonlinear activation calibrate the means and STDs.
The training progresses in the following order: we implement several trials as a mini-batch, utilize speaker embedding to get the scores, and compute the threshold within the mini-batch. 
We subtract the threshold value for each obtained score and then apply the sigmoid activation function.
Through this process, the threshold value becomes $0.5$ for every mini-batch, which allows for optimization in normalizing the scores.
The negative log-likelihood loss is used to learn to be similar to the target values consisting of 1 or 0.
Specifically, we used the dataset that used for fine-tuning in Section 3.3.
We configured a mini-batch similar to the challenge scenario to implement the robust TAS-Norm for the challenge.
No data augmentation was used for the enrollment utterances, and the length of the utterances varied from 1 s to 1.8 s for the test utterances.
\subsection{Quality-aware score calibration}
To boost the SV performance, quality-aware score calibration \cite{id_voxsrc20} was employed with validation trials, where the score was calibrated by assigning a value to each trial of quality measuring functions (QMFs).
We constructed subsets from the MLS and ROBOVOX samples of 175 and 2 speakers, respectively, to construct the validation trials.
We randomly selected 100 utterances for each speaker, for a total of 17,700 utterances for a subset.
Using these utterances, we constructed 150,000 trials, where, similar to the challenge environment, enrollment utterances were not augmented, and test utterances were augmented as described in Section 3.4, depending on the type of dataset.
For the ROBOVOX sample dataset, audio from 4th and 5th channel of the multi-channel sample was used only for enrollment and test utterances, respectively.
We set the ratio of imposter to target as 10 to 1.
The calibrated score was obtained according to the following equation:
\begin{equation}
\label{eqn:qmf}
    S_{fusion} = \left[x_{1} \dots x_{k}\right]
    \begin{bmatrix}
        S(1) \\
        \dots \\
        S(k)
    \end{bmatrix}
    +
    \left[y_{1} \dots y_{n}\right]
    \begin{bmatrix}
        Q(1) \\
        \dots \\
        Q(n)
    \end{bmatrix}
    ,
\end{equation}
where $x_{k}$ is a $k$-th score's weight, $S(k)$ is a $k$-th system's normalized scores,  $y_{n}$ is a $n$-th QMF weight and $Q(n)$ is a $n$-th QMF value.
To find the score and QMF weights, we used the Logistic Regression to map the output score $S_{fusion}$ as range of 0 to 1.
QMF is an auxiliary measure that is extracted for the input signal or speaker embedding. 
It is assumed to provide additional information that is not captured by the single model utterance embedding \cite{ID_voxsrc23}.
Therefore, we utilized the supplemental information mentioned in \cite{ID_voxsrc23} for the test trials.
\begin{itemize}[leftmargin=0.25in]
\item Audio quality measurements (AQMs): given that reverberation and noise prevented the VADs from working properly, we used the four aforementioned VADs as QMFs.
\item Speaker embedding metrics (SEMs): we extracted the speaker embeddings for each utterance, computed the statistics of the features, and used them as QMFs. We used the $l_{1}$ and $l_{2}$ norms of the embedding and STD for the embedding dimension. The mean and STD calculated from the AS-Norm, and the mean and STD calibrated from the TAS-Norm were used as QMFs.
\end{itemize}
\section{Evaluation result}
Table 8 shows the evaluation results of single and fusion systems on the ROBOVOX multi-channel and single-channel trials.
For the ECAPA-TDNN-2048, the results of No.\,1 through 4 used the settings of No.\,6 in Table 6, and the experimental settings of No.\,5 were identical to the experimental settings of No.\,10 in Table 6.
In the ResNet-64 results, No.\,6 through 9 used the settings of No.\,8 in Table 7, and No.\,10 used settings of No.\,10 in Table 7.
From the results in Table 8 we can see that the proposed TAS-Norm was more effective than the AS-Norm.
This confirms that the TAS-norm compensated for the deficiencies of the AS-Norm in calibrating the score.
And the dereverberated signals through the VACE were not very effective in this challenge, which can be attributed to the fact that the pre-trained VACE model was immediately imported.
We believe that additional fine-tuning to optimize jointly with the speaker embedding model can solve this issue of domain mismatch.
Finally, we confirmed the effectiveness of using QMF in the fusion process.
SEMs improved the SV performance a lot, while AQMs slightly improved the SV performance.
The final submitted system was in No.\,14, which used both QMF methods.
\section{Conclusions}
In this report, we detailed our system for the far-field single-channel track of the ROBOVOX challenge.
We employed the ResNet-based speaker embedding model and ECAPA-TDNN with margin softmax-based loss functions.
We have found the significant importance of the data setup for both training and testing and the usage of QMF values in fusion.
We also developed the TAS-Norm and trained it for considering the challenge condition to improve SV performance.
In future work, we would like to add more experiments on the proposed TAS-Norm to observe if it can learn robustly to the evaluation environment.
Our submission achieved second place on the SP Cup 2024 public leaderboard \footnote{https://www.codabench.org/competitions/1588}.
After, we placed second place on the finals on April 15, 2024.

%
\section{Acknowledgments}
This work was supported by Samsung Research.

\vfill\pagebreak
\bibliographystyle{IEEEbib}
\bibliography{refs}

\end{document}

%% file: tables/data_stats.tex
\begin{table}[t] \begin{center}
\caption{Statistics of training datasets.}
\resizebox{\columnwidth}{!}{
\begin{tabular} {cccc}
\toprule[1.2pt]
\textbf{Dataset} & \textbf{Language} & \textbf{\# of speakers} & \textbf{\# of audio sample} \\
\midrule
VoxCeleb2 $\textit{dev}$ & Multilingual & 5,994 & 1,092,009 \\
CN-Celeb1 & Chinese & 997 & 126,532 \\
SdSV2020 Task 2 & Persian & 588 & 85,764 \\
mTEDx (Fr) & French & 961 & 83,879 \\
MLS (Fr) & French & 145 & 53,070 \\
ROBOVOX sample & French & 2 & 147 \\
\midrule
Total & & 8,687 & 1,441,401 \\
\bottomrule[1.2pt]
\end{tabular}
}
\end{center}
\end{table}

%% file: tables/ecapa-tdnn-c.tex
\begin{table}[t] \begin{center}
\caption{ECAPA-TDNN-C architecture} 
\vspace{0.2cm}
\begin{tabular}{lcr}
\toprule[1.2pt]
\textbf{Layer name} & \textbf{Structure} & \textbf{Output} \\
\midrule
First-Conv1D & 5, C, stride=1 & $\mathrm{C \times T}$ \\
\midrule
\multirow{4}{*}{SE-Res2Block-1} & Res2Conv1D & \multirow{2}{*}{$\mathrm{C \times T}$}  \\
    & 3, C, dilation=2, scale=8 &  \\
    & SE-Block & \multirow{2}{*}{$\mathrm{C \times T}$} \\
    & bottleneck=256 & \\
\midrule
\multirow{4}{*}{SE-Res2Block-2} & Res2Conv1D & \multirow{2}{*}{$\mathrm{C \times T}$}  \\
    & 3, C, dilation=3, scale=8 &  \\
    & SE-Block & \multirow{2}{*}{$\mathrm{C \times T}$} \\
    & bottleneck=256 & \\
\midrule
\multirow{4}{*}{SE-Res2Block-3} & Res2Conv1D & \multirow{2}{*}{$\mathrm{C \times T}$}  \\
    & 3, C, dilation=4, scale=8 &  \\
    & SE-Block & \multirow{2}{*}{$\mathrm{C \times T}$} \\
    & bottleneck=256 & \\
\midrule
Last-Conv1D & 1, 1536, stride=1 & $\mathrm{1536 \times T}$ \\
\midrule
Pooling & CD-ASP & $\mathrm{3072 \times 1}$ \\
\midrule
Dense & FC & $\mathrm{256 \times 1}$ \\
\bottomrule[1.2pt]
\end{tabular}
\end{center}
\end{table}

%% file: tables/exp_duration.tex
\begin{table}[t] \begin{center}
\caption{Experimental result of input audio duration using the ECAPA-TDNN-512}
\vspace{0.2cm}
\resizebox{0.9\columnwidth}{!}{
\begin{tabular} {cccccc}
\toprule[1.2pt]

\multicolumn{1}{c}{\multirow{2}{*}{\textbf{No.}}}&
\multicolumn{2}{c}{\textbf{Duration}}&
\multicolumn{3}{c}{\textbf{ROBOVOX Multi-channel trial}}\\
& {Speech} & {Noise} && {EER (\%)} & {$\text{DCF}_{c}$}  \\
\midrule
{1.} & {2.0 s} & {2.0 s} && {16.03} & {0.7526} \\
{2.} & {2.0 s} & {3.0 s} && {15.09} & {0.7451} \\
{3.} & {2.0 s} & {0.5 s} && {17.02} & {0.7701} \\
{4.} & {1.5 s} & {2.5 s} && {16.63} & {0.7657} \\
{5.} & {1.5 s} & {2.5 s} && {15.67} & {0.7729} \\
{6.} & {1.0 s} & {2.0 s} && {15.71} & {0.7619} \\
{7.} & {1.8 s} & {2.4 s} && \textbf{14.98} & \textbf{0.7432} \\
\bottomrule[1.2pt]
\end{tabular}
}
\end{center}
\end{table}

%% file: tables/exp_augmentation.tex
\begin{table}[t] \begin{center}
\caption{Ablation study of data augmentation parameter using the ECAPA-TDNN-512}
\vspace{-0.2cm}
\resizebox{\columnwidth}{!}{
\begin{tabular} {ccccccc}
\toprule[1.2pt]

\multicolumn{1}{c}{\multirow{2}{*}{\textbf{No.}}}&
\multicolumn{3}{c}{\textbf{Augmentation}}&
\multicolumn{3}{c}{\textbf{ROBOVOX Multi-channel trial}}\\
& {RIR Prob.} & {MUSAN} & {Clip Prob.} && {EER (\%)} & {$\text{DCF}_{c}$}  \\
\midrule
{1.} & {0.5} & & && {15.09} & {0.7451} \\
{2.} & {0.5} & {- music}& && {14.75} & {0.7264} \\
{3.} & {0.25} & {- music}& && {15.04} & {0.7380} \\
{4.} & {0.75} & {- music}& && {14.98} & {0.7365} \\
{5.} & {0.5} & {- music}& {0.5} && {14.82} & {0.7315} \\
{6.} & {0.5} & {- music}& {0.25} && \textbf{14.50} & \textbf{0.7174} \\
\bottomrule[1.2pt]
\end{tabular}
}
\end{center}
\end{table}

%% file: tables/resnet-64.tex
\begin{table}[t] \begin{center}
\caption{ResNet-64 architecture}
\vspace{-0.2cm}
\centering
\resizebox{\columnwidth}{!}{
\begin{tabular}{lcr}
\toprule[1.2pt]
\textbf{Layer name} & \textbf{Structure} & \textbf{Output} \\ 
\midrule
First-Conv2D & $3 \times 3$, 64, stride=1 & $\mathrm{64 \times 96 \times T}$  \\
\midrule
ResBlock-1 & $\begin{bmatrix}
    3 \times 3, 64 \\
    3 \times 3, 64 \\
    \mathrm{fwSE},[128, 96]
\end{bmatrix} \times 3$ & $\mathrm{64 \times 96 \times T}$  \\
\midrule
ResBlock-2 & $\begin{bmatrix}
    3 \times 3, 128 \\
    3 \times 3, 128 \\
    \mathrm{fwSE},[128, 48]
\end{bmatrix} \times 4$ & $\mathrm{128 \times 48 \times T/2}$ \\
\midrule
ResBlock-3 & $\begin{bmatrix}
    3 \times 3, 256 \\
    3 \times 3, 256 \\
    \mathrm{fwSE},[128, 24]
\end{bmatrix} \times 21$ & $\mathrm{256 \times 24 \times T/4}$ \\
\midrule
ResBlock-4 & $\begin{bmatrix}
    3 \times 3, 256 \\
    3 \times 3, 256 \\
    \mathrm{fwSE},[128, 12]
\end{bmatrix} \times 3$ & $\mathrm{256 \times 12 \times T/8}$ \\
\midrule
Flatten(C, F) & --- & $\mathrm{3072 \times T/8}$ \\
\multirow{2}{*}{Pooling} & CD-ASP & $\mathrm{6144 \times 1}$ \\
& FS-ASP & $\mathrm{1024 \times 1}$ \\
Concatenate & --- & $\mathrm{7168 \times 1}$ \\
\midrule
Dense & FC & $\mathrm{256 \times 1}$ \\
\bottomrule[1.2pt]
\end{tabular}
}
\end{center}
\end{table}
%

%

%% file: fig1.tex
\begin{figure}[t]
  \centering
  \includegraphics[width=\linewidth]{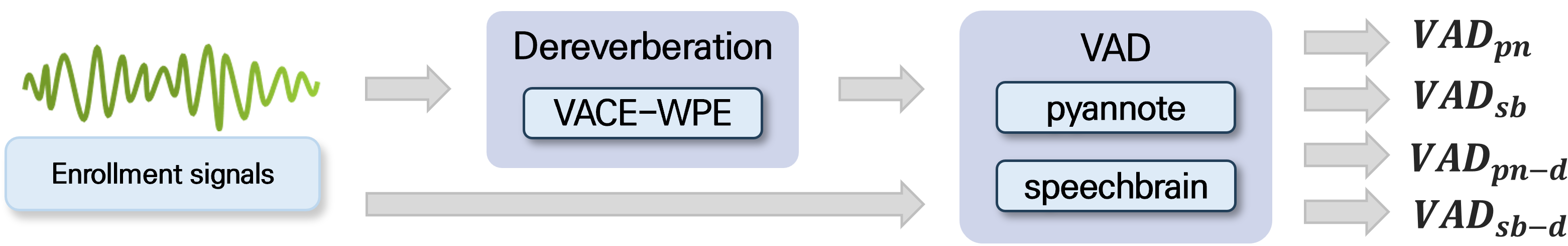}
  \vspace{-0.1cm}
  \caption{Block diagram of overall pre-processing process.}
  \label{figure1}
  \vspace{-0.1cm}
\end{figure}

%% file: tables/exp_preprocessiong1.tex
\begin{table}[t] \begin{center}
\caption{Ablation study of pre-processing using the ECAPA-TDNN-2048. For simplicity, we denote the original signal as \textit{org} and dereverberated signal as \textit{der}.}
\vspace{0.2cm}
\resizebox{\columnwidth}{!}{
\begin{tabular} {ccccccc}
\toprule[1.2pt]

\multicolumn{1}{c}{\multirow{2}{*}{\textbf{No.}}}&
\multicolumn{1}{c}{\multirow{2}{*}{\textbf{Data}}}&
\multicolumn{2}{c}{\textbf{Pre-processing}}&
\multicolumn{3}{c}{\textbf{ROBOVOX Multi-channel trial}}\\
& & {VAD} & {Segmentation} && {EER (\%)} & {$\text{DCF}_{c}$}  \\
\midrule
{1.} & {\textit{org}} & VAD$_{pn}$ & && {10.25} & {0.5900} \\
{2.} & {\textit{org}} & VAD$_{pn}$ & {\checkmark} && {8.40} & {0.5173} \\
{3.} & {\textit{org}} & VAD$_{sb}$ & && {10.09} & {0.5582} \\
{4.} & {\textit{org}} & VAD$_{sb}$ & {\checkmark} && {9.20} & \textbf{0.5157} \\
{5.} & {\textit{org}} & VAD$_{pn-d}$ & && {10.01} & {0.5695} \\
{6.} & {\textit{org}} & VAD$_{pn-d}$ & {\checkmark} && \textbf{7.94} & {0.5308} \\
{7.} & {\textit{org}} & VAD$_{sb-d}$ & && {10.43} & {0.5816} \\
{8.} & {\textit{org}} & VAD$_{sb-d}$ & {\checkmark} && {8.99} & {0.5179} \\
\midrule
{9.} & {\textit{der}} & VAD$_{pn-d}$ & && {13.14} & {0.6538} \\
{10.} & {\textit{der}} & VAD$_{pn-d}$ & {\checkmark} && \textbf{8.41} & {0.6073} \\
{11.} & {\textit{der}} & VAD$_{sb-d}$ & && {14.13} & {0.6661} \\
{12.} & {\textit{der}} & VAD$_{sb-d}$ & {\checkmark} && {9.91} & \textbf{0.5974} \\
\bottomrule[1.2pt]
\end{tabular}
}
\end{center}
\end{table}

%% file: tables/exp_preprocessiong2.tex
\begin{table}[t] \begin{center}
\caption{Ablation study of pre-processing using the ResNet-64. For simplicity, we denote the original signal as \textit{org} and dereverberated signal as \textit{der}.}
\vspace{0.2cm}
\resizebox{\columnwidth}{!}{
\begin{tabular} {ccccccc}
\toprule[1.2pt]

\multicolumn{1}{c}{\multirow{2}{*}{\textbf{No.}}}&
\multicolumn{1}{c}{\multirow{2}{*}{\textbf{Data}}}&
\multicolumn{2}{c}{\textbf{Pre-processing}}&
\multicolumn{3}{c}{\textbf{ROBOVOX Multi-channel trial}}\\
& & {VAD} & {Segmentation} && {EER (\%)} & {$\text{DCF}_{c}$}  \\
\midrule
{1.} & {\textit{org}} & VAD$_{pn}$ & && {10.16} & {0.5635} \\
{2.} & {\textit{org}} & VAD$_{pn}$ & {\checkmark} && {9.52} & \textbf{0.5540} \\
{3.} & {\textit{org}} & VAD$_{sb}$ & && {10.47} & {0.5657} \\
{4.} & {\textit{org}} & VAD$_{sb}$ & {\checkmark} && {10.17} & {0.5553} \\
{5.} & {\textit{org}} & VAD$_{pn-d}$ & && {10.61} & {0.5729} \\
{5.} & {\textit{org}} & VAD$_{pn-d}$ & {\checkmark} && {9.80} & {0.5824} \\
{7.} & {\textit{org}} & VAD$_{sb-d}$ & && {10.37} & {0.5611} \\
{8.} & {\textit{org}} & VAD$_{sb-d}$ & {\checkmark} && \textbf{9.51} & {0.5664} \\
\midrule
{9.} & {\textit{der}} & VAD$_{pn-d}$ & && {11.82} & {0.6074} \\
{10.} & {\textit{der}} & VAD$_{pn-d}$ & {\checkmark} && \textbf{10.37} & \textbf{0.5611} \\
{11.} & {\textit{der}} & VAD$_{sb-d}$ & && {12.52} & {0.6163} \\
{12.} & {\textit{der}} & VAD$_{sb-d}$ & {\checkmark} && {10.47} & {0.5657} \\
\bottomrule[1.2pt]
\end{tabular}
}
\end{center}
\end{table}

%% file: tables/exp_final.tex
\begin{table*}[ht] \begin{center}
\caption{Evaluation results on the ROBOVOX multi-channel and single-channel trials.}
\vspace{0.3cm}
\resizebox{\textwidth}{!}{
\begin{tabular} {ccccccccccc}
\toprule[1.2pt]
\multicolumn{1}{c}{\multirow{2}{*}{\textbf{No.}}}&
\multicolumn{1}{c}{\multirow{2}{*}{\textbf{Models}}}&
\multicolumn{1}{c}{\multirow{2}{*}{\textbf{Step / Option}}}&
\textbf{Score}&
\textbf{Dereverberated}&
\multicolumn{3}{c}{\textbf{ROBOVOX Multi-channel trial}}&
\multicolumn{3}{c}{\textbf{ROBOVOX Single-channel trial}}\\ 
&&& \textbf{calibration} & \textbf{signal} && {EER (\%)} & {$\text{DCF}_{c}$} && {EER (\%)} & {$\text{DCF}_{c}$} \\
\midrule
{1.} & \multicolumn{1}{c}{\multirow{5}{*}{ECAPA-TDNN-2048}} &
{pre-training} & & 
&& {10.81} & {0.5549} && {11.26} & {0.7728}\\
\cmidrule(l){3-11}
{2.} & &
\multicolumn{1}{c}{\multirow{4}{*}{fine-tuning}} & & 
&& {7.94} & {0.5308} && {8.20} & {0.7020}\\
{3.} & &
 & {+ AS-Norm} & 
&& {7.81} & {0.4821} && {8.02} & {0.7046}\\
{4.} & &
 & {+ TAS-Norm} & 
&& \textbf{7.38} & \textbf{0.4690} && \textbf{7.82} & {0.6785}\\
{5.} & &
 & {+ TAS-Norm} & {\checkmark}
&& {8.05} & {0.5658} && {8.45} & \textbf{0.6383}\\
\midrule
{6.} & \multicolumn{1}{c}{\multirow{5}{*}{ResNet-64}} &
{pre-training} & & 
&& {10.39} & {0.6050} && {12.15} & {0.8025}\\
\cmidrule(l){3-11}
{7.} & &
\multicolumn{1}{c}{\multirow{4}{*}{fine-tuning}} & & 
&& {9.51} & {0.5664} && {10.10} & {0.7414}\\
{8.} & &
 & {+ AS-Norm} & 
&& {9.16} & \textbf{0.5407} && {9.79} & {0.7348}\\
{9.} & &
 & {+ TAS-Norm} & 
&& \textbf{8.54} & {0.5450} && {9.64} & {0.7356}\\
{10.} & &
 & {+ TAS-Norm} & {\checkmark}
&& {9.24} & {0.5933} && \textbf{9.59} & \textbf{0.7322}\\
\midrule
{11.} & \multicolumn{1}{c}{\multirow{8}{*}{Fusion}} & \multicolumn{1}{c}{\multirow{4}{*}{No. (4,9)}}
& without QMF & && {6.89} & {0.4684} && {7.76} & {0.6789}\\
{12.} &  & 
& + QMF (AQM) & && {6.64} & {0.4542} && {8.55} & {0.5946}\\
{13.} &  & 
& + QMF (SEM) & && {6.48} & {0.3988} && \textbf{6.13} & {0.6125}\\
{14.} &  & 
& + QMF (use both) & && \textbf{5.91} & \textbf{0.3865} && {6.46} & \textbf{0.5245}\\
\cmidrule(l){3-11}
{15.} &  & \multicolumn{1}{c}{\multirow{4}{*}{No. (4,5,9,10)}}
& without QMF & && {6.95} & {0.4708} && {7.84} & {0.6751}\\
{16.} &  & 
& + QMF (AQM) & && {6.68} & {0.4226} && {8.10} & \textbf{0.5723}\\
{17.} &  & 
& + QMF (SEM) & && {5.23} & \textbf{0.3789} && \textbf{5.68} & {0.6041}\\
{18.} &  & 
& + QMF (use both) & && \textbf{5.22} & {0.3798} && {5.88} & {0.6066}\\
\bottomrule[1.2pt]
\end{tabular}}
\end{center}
\vspace{-0.3cm}
\end{table*}